\documentclass[sigconf]{acmart}

\copyrightyear{2024}
\acmYear{2024} 
\setcopyright{iw3c2w3}
\acmConference[WWW '24]{Proceedings of the Web Conference 2024}{May 13-17, 2024}{Singapore} 
\acmPrice{}
\acmDOI{}
\acmISBN{}
\usepackage{microtype}
\usepackage{graphicx}
\usepackage{subfigure}
\usepackage{booktabs} 

\usepackage{algorithm}
\usepackage{algorithmic}
\usepackage{tabularx}
\usepackage{adjustbox}
\usepackage{hyperref}

\usepackage{amsmath}
\usepackage{mathtools}
\usepackage{amsthm}
\usepackage{amsfonts}       
\usepackage[utf8]{inputenc} 
\usepackage[T1]{fontenc}    
\usepackage{xurl}
\usepackage{nicefrac}       
\usepackage{xcolor}         
\usepackage{bbm}
\usepackage[utf8]{inputenc}
\usepackage[T1]{fontenc}    
\usepackage{xcolor}         
\usepackage{natbib}
\usepackage{xspace}
\usepackage{enumerate, paralist}
\usepackage{natbib}
\usepackage{comment}
\usepackage{multirow}
\usepackage{xspace}
\usepackage{rotating}
\usepackage{graphicx}
\usepackage{booktabs} 
\usepackage{tabularx}
\usepackage{enumitem}
\usepackage{thmtools} 
\usepackage{thm-restate}
\theoremstyle{plain}

\theoremstyle{definition}

\begin{document}

\title{Neural Contextual Bandits for Personalized  Recommendation}

\author{Yikun Ban}
\affiliation{%
  \institution{University of Illinois at Urbana-Champaign}
  \country{}
} 
\email{yikunb2@illinois.edu}

\author{Yunzhe Qi}
\affiliation{%
  \institution{University of Illinois at Urbana-Champaign}
  \country{}
 }
\email{yunzheq2@illinois.edu}

\author{Jingrui He}
\affiliation{%
  \institution{University of Illinois at Urbana-Champaign}
  \country{}
}
\email{jingrui@illinois.edu}

\begin{abstract}

In the dynamic landscape of online businesses, recommender systems are pivotal in enhancing user experiences. While traditional approaches have relied on static supervised learning, the quest for adaptive, user-centric recommendations has led to the emergence of the formulation of contextual bandits. This tutorial investigates the contextual bandits as a powerful framework for personalized recommendations. We delve into the challenges, advanced algorithms and theories, collaborative strategies, and open challenges and future prospects within this field.
Different from existing related tutorials, (1) we focus on the exploration perspective of contextual bandits to alleviate the ``Matthew Effect'' in the recommender systems, i.e., the rich get richer and the
poor get poorer, concerning the popularity of items; (2) in addition to the conventional linear contextual bandits, we will also dedicated to neural contextual bandits which have emerged as an important branch in recent years, to investigate how neural networks benefit contextual bandits for personalized recommendation both empirically and theoretically; (3) we will cover the latest topic,   collaborative neural contextual bandits, to incorporate both user heterogeneity and user correlations customized for recommender system; (4) we will provide and discuss the new emerging challenges and open questions for neural contextual bandits with applications in the personalized recommendation, especially for large neural models.   

Compared with other greedy personalized recommendation approaches, Contextual Bandits techniques provide distinct ways of modeling user preferences.
We believe this tutorial can benefit researchers and practitioners by appreciating the power of exploration and the performance guarantee brought by neural contextual bandits, as well as rethinking the challenges caused by the increasing complexity of neural models and the magnitude of data.

\end{abstract}

\maketitle

\section{Introduction}

Recommender systems are indispensable in various online businesses, including e-commerce platforms and online streaming services. They leverage user correlations to assist the perception of user preferences, a field of study spanning several decades. In the past, considerable effort has been directed toward supervised-learning-based collaborative filtering methods within relatively static environments \citep{su2009survey, he2017neural}.
However, the ideal recommender systems should adapt over time to consistently meet user interests. Consequently, it is natural to formulate the recommendation process as a sequential decision-making process. In this paradigm, the recommender engages with users, observes their online feedback (i.e., rewards), and optimizes the user experience for long-term benefits, rather than fitting a model on the collected static data based on supervised learning \cite{xue2023resact,chen2022off,gao2023alleviating}.
Based on this idea, this tutorial focuses on the formulation of contextual bandits \cite{2011improved,2014onlinecluster,2016collaborative,gentile2017context,2019improved,ban2021local, qi2022neural,mcdonald2023impatient,qi2023graph, ban2022improved, ban2020generic}.
In particular, the emerging neural bandit techniques \cite{zhou2020neural,zhang2020neural,ban2021multi,jia2021learning,ban2021ee,dai2022federated,kassraie2022neural, qi2023meta, ban2021convolutional} have shown their superiority over linear or kernel \cite{auer2002finite,2011improved,li2010exploitation,valko2013finite,deshmukh2017multi} methods due to the representation power of neural networks. As a result, they have been considered the ideal solution to deal with the non-linear problem settings, in terms of the Contextual Bandits research and application.

To be specific, let us dive into a scenario involving a total of $T$ rounds of recommendations. 
In the $t$-th round, the learner (model) is presented with $K$ arms (items), each of which yields a reward based on an unknown function that captures the users' preferences.  
The learner's task is to select one arm in each round and recommend this arm to the serving user. Consequently, the learner observes the resulting rewards and updates its recommendation policy. The ultimate goal is to maximize the cumulative rewards across the $T$ rounds, i.e., to minimize the cumulative regrets incurred over the course of these interactions.
However, the fundamental challenge of balancing exploitation and exploration inherently arises in the context of sequential recommendation. In a round, the learner faces a dilemma. On one hand, it must `exploit' the knowledge gleaned from the previous rounds to prioritize popular items. On the other hand, the model is also tasked with `exploring' potential value from new or under-explored items in seek of long-term benefits.
As a result, striking the right balance between exploitation and exploration is imperative for constructing a robust recommender system and contextual bandits provide powerful tools to tackle this dilemma.

It is well known that research works on "User Modeling and Recommendation" have always been an indispensable part of the Web Conference. 
Different from other greedy personalized recommendation approaches (e.g., the conventional collaborative filtering methods \citep{su2009survey, he2017neural}), Contextual Bandits techniques provide distinct ways of modeling user preferences \cite{2014onlinecluster,2016collaborative,gentile2017context,2019improved,ban2021local,korda2016distributed,yang2020exploring,wu2021clustering}, by balancing exploiting the process knowledge and exploring the unrevealed benefits.
Therefore, all of these aspects make this tutorial a good fit for the Web Conference and its community members.

In this lecture-style tutorial, we will provide a comprehensive review of contextual bandits algorithms under the personalized recommendation settings.
We will begin by delving into the challenges of personalized recommendation and why contextual bandits emerge as a powerful tool.
The subsequent section will provide an in-depth review of advanced algorithms and theories in contextual bandits, ranging from linear to neural contextual bandits, emphasizing their role in enhancing exploration in personalized recommendation.
The third part is the collaborative contextual bandits, extending beyond individual bandits to account for user correlations. We will explore two intriguing problem formulations: clustering of bandits, which is to formulate the scenarios in which we leverage the knowledge of a cluster of users to facilitate the decision-making of the serving user, and graph bandits learning, which is to formulate fine-grained user correlations represented by the strength of edges in the constructed graph.
In conclusion, we will shed light on the open challenges and future directions in contextual bandits, particularly within personalized recommendation scenarios. Key areas of focus will include the scalability of contextual bandits in large recommender systems and ensuring the trustworthiness of personalized decision-making.
This tutorial promises to be a captivating review and lecture on the evolving contextual bandits that holds immense potential for revolutionizing how we perform content recommendation to users.

\section{Target Audience}
The tutorial is developed for all the people who are interested in the related research areas, e.g., multi-armed bandits, reinforcement learning, information retrieval, data mining,  recommender systems, etc. 
The audience is expected to have basic knowledge of machine learning and data mining. 
This tutorial balances the introductory and advanced material, i.e., 50 \% for beginners and 50 \% for intermediate and advanced researchers.

In addition, as this tutorial will be an organic combination of (1) the application of bandit techniques in recommendation, as well as (2) the theoretical insights of bandit algorithms, we believe that both the business practitioner as well as researchers focusing on algorithmic aspects will benefit from our contents.

\section{Short Bio}

In this section, we briefly introduce the three presenters of this tutorial.

\noindent \textbf{Yikun Ban} is a final-year Ph.D. student in the Department of Computer Science at the University of Illinois at Urbana–Champaign. He is a member of DAIS (Data and Information Systems) Research Lab. He received his M.CS. degree from Peking University in 2019 and B.Eng. degree from Wuhan University in 2016. His research interests lie in multi-armed bandits/Reinforcement Learning to design and develop principled exploration strategies in sequential decision-making. He has published more than 11 papers at top conferences in Machine Learning and Data Mining (e.g., WWW, KDD, NeurIPS, ICLR, AAAI) and has been a reviewer or program committee member of mainstream machine learning journals and conferences. He was an applied scientist intern at Amazon Web Service, and his research works have been powering primary applications in Amazon and Instacart.

\noindent \textbf{Yunzhe Qi} is a Ph.D. candidate in the School of Information Sciences at the University of Illinois at Urbana–Champaign (UIUC). He received his Master's degree from UIUC in 2021, and his Bachelor's degree in Beijing University of Posts and Telecommunications in 2019 respectively. His research interests mainly focus on utilizing Contextual Bandit methods to solve the exploitation-exploration dilemma for machine learning tasks, such as online recommendation.
He has published several papers at top machine learning / data mining conferences (e.g., KDD, NeurIPS), and has been serving as the reviewer as well as PC member for multiple machine learning / data mining conferences and journals. He was a Machine Learning Engineer Intern at Instacart, who designed and implemented Contextual Bandit frameworks for personalized recommendation that have been generating actual business growth.

\noindent \textbf{Jingrui He} (Corresponding Tutor) is a professor in the School of Information Sciences at the University of Illinois Urbana-Champaign. She received her Ph.D. from Carnegie Mellon University in 2010. Her research focuses on heterogeneous machine learning, rare category analysis, active learning and semi-supervised learning, with applications in security, social network analysis, healthcare, and manufacturing processes. Dr. He is the recipient of the 2016 NSF CAREER Award, the 2020 OAT Award, three-time recipient of the IBM Faculty Award in 2018, 2015, and 2014, and was selected as IJCAI 2017 Early Career Spotlight. Dr. He has more than 100 publications at major conferences (e.g., WWW, IJCAI, AAAI, KDD, ICML, NeurIPS) and journals (e.g., TKDE, TKDD, DMKD), and is the author of two books. Her papers have received the Distinguished Paper Award at FAccT 2022, as well as Bests of the Conference at ICDM 2016, ICDM 2010, and SDM 2010. She has several years of course teaching experience as an instructor and has offered several tutorials at major conferences, e.g., KDD, AAAI, IJCAI, SDM, IEEE BigData, etc. in the past few years. For more information, please refer to her homepage at https://www.hejingrui.org/.

\section{Outline}

This will be a  \textbf{3-hour lecture-style} tutorial to cover the state-of-the-art research for neural contextual bandit algorithms and theories with applications in personalized recommendation.

\begin{itemize}
    \item \textbf{Introduction} \\
        \textit{Duration:} 15 minutes,    \textit{Presenter:} Yikun Ban \\
        We start by motivating the formulation of contextual bandits for personalized recommendation and its existing challenges.
    \begin{itemize}
        \item Motivation
        \item Formulation of Sequential Decision-Making \cite{slivkins2019introduction}
        \item Challenges
    \end{itemize}

 \item \textbf{Part I: Linear Contextual Bandits} \\
      \textit{Duration:} 30 minutes,    \textit{Presenter:} Yikun Ban \\
      In this part, we introduce the problem definition of linear contextual bandits and the existing linear exploration strategy with applications in personalized recommendation.
      \begin{itemize}
        \item Linear Upper Confidence Bound (UCB) \cite{2011improved}
        \item Linear Thompson Sampling (TS) \cite{agrawal2013thompson}
        \item Linear Personalized Recommendation \cite{2010contextual}
    \end{itemize}
 \item \textbf{Part II: Neural Contextual Bandits} \\
 Duration: 40 minutes,    Presenter: Yunzhe Qi and Yikun Ban \\
 This is the core part. We introduce the problem definition, latest algorithms, and theoretical guarantee for neural contextual bandits, with elaboration on how the neural networks benefit contextual bandits for personalized recommendation.
    \begin{itemize}
        \item Neural Upper Confidence Bound (UCB) \cite{zhou2020neural}
        \item Neural Thompson Sampling (TS) \cite{zhang2020neural}
        \item Exploitation-Exploration Networks \cite{ban2021ee, ban2023neural}
        \item Neural Inverse Gap Weighting \cite{foster2020beyond, deb2023contextual}
        \item Neural Personalized Recommendation
    \end{itemize}
\item \textbf{Part III Collaborative Contextual Bandits} \\
\textit{Duration:} 60 minutes,    \textit{Presenter:} Yunzhe Qi and Yikun Ban \\
This is the core part. We introduce the motivation of collaborative Contextual Bandits and focus on two problems: clustering of bandits and graph bandits learning. 
\begin{itemize}
    \item Motivation and Challenges
    \item Clustering of Bandits
    \begin{itemize}
        \item Linear Clustering of Bandits \cite{ban2021local, gentile2017context}
        \item Neural Clustering of Bandits  \cite{ban2022neural}
    \end{itemize}
    \item Graph Bandits Learning \cite{qi2023graph, qi2022neural}
    \item Applications in Recommender Systems \cite{ban2021multi}
\end{itemize}

\item \textbf{Part IV: Open Questions and Future Trends} \\
\textit{Duration:} 35 minutes,    \textit{Presenter:} Yikun Ban and Jingrui He \\
In this part, we will discuss the emerging challenges and open questions for neural contextual bandits in applications of recommender systems.
\begin{itemize}
    \item Large Search Space: Arm and User Space
    \item Transparency: Rationales and Explanation for Exploration
    \item Fairness: Exploit-Explore Fairly  
    \item Privacy: Privacy-preserving Decision-making 
\end{itemize}
\end{itemize}

\section{Related Tutorials or Talks}


\begin{itemize}
\item \textit{Practical Bandits: An Industry Perspective} \\
The WebConf 2023 Tutorial  \\
\textbf{Event date:} May 3, 2023.     \textbf{Location:} Austin, Texas, USA \\
\textbf{Differences:} In this WebConf tutorial, they present bandits algorithms from a practical perspective to facilitate practitioners to determine non-contextual or contextual approaches, on- or off-policy setups, delayed or immediate feedback, etc. Instead, our tutorial focuses on the neural contextual bandits that have attained increasing attention recently year from both practical and theoretical aspects, and our backbone is for SOTA exploration strategies with applications in the recommender system.

\item \textit{Bridging Learning and Decision Making} \\
The ICML 2022 Tutorial \\
\textbf{Event Date:}  July 18, 2022.         \textbf{Location:} Baltimore, Maryland, USA \\
\textbf{Difference:}
This ICML tutorial gives an overview of the theoretical foundations of contextual bandits and reinforcement learning. In particular, they focus on the unified bandit framework to build the connection between supervised estimation and sequential decision-making.  However, they lack the coverage of neural contextual bandits and collaborative contextual bandits tailored for the recommender system, which is the main gap we try to fill in this tutorial.    
\end{itemize}

\section{Previous Editions}
This will be the first edition of this tutorial, but the presenters have experience in teaching material covering similar topics in the past. We anticipate to present (an extended version of) this tutorial at similar conferences in the future.

\section{Tutorial materials}
We will set up a website to release all the related materials, including presentation slides, references, and open-source data \& code, in order to get the audience familiarized with the tutorial content before the tutorial begins. Besides, all the researchers and practitioners are encouraged to have Q\&A interaction during the tutorial or further discussions offline.

\section{Link to Video Teaser}
The video teaser file can be found at the following Dropbox link: \url{https://www.dropbox.com/scl/fo/32idrzywjz6gc9qp3iixi/h?rlkey=0xagsm4m4hhfb2vzxg6hx8bjv&dl=0}

\bibliographystyle{abbrvnat}
\bibliography{ref}

\end{document}